\documentclass[showpacs,preprintnumbers,amssymb] {revtex4}
\begin{document}
\def\baselinestretch{1.}

\title{New exact solutions of the standard pairing model for well-deformed nuclei}
\author{Feng Pan,$^{a,b}$,  Ming-Xia Xie,$^{a}$ Xin
Guan,$^{a}$ Lian-Rong Dai,$^{a}$ and J. P. Draayer$^{b}$}
\address{ $^{a}$Department of Physics, Liaoning Normal University,
Dalian, 116029, P. R. China\\
$^{b}$Department of Physics and Astronomy, Louisiana State
University, Baton Rouge, LA 70803,USA}

\vspace{0.3cm}

\vskip .5cm \begin{abstract} A new step-by-step diagonalization
procedure for evaluating exact solutions of the nuclear deformed
mean-field plus pairing interaction model is proposed via a simple
Bethe ansatz in each step from which the eigenvalues and
corresponding eigenstates can be obtained progressively. This new
approach draws upon an observation that the original one- plus
two-body problem in a $k$-particle Hilbert subspace can be mapped
unto a one-body grand hard-core boson picture that can be solved
step by step with a simple Bethe ansatz known from earlier work.
Based on this new procedure, it is further shown that the extended
pairing model for deformed nuclei [Phys. Rev. Lett. {\bf 92}, 112503
(2004) ] is similar to the standard pairing model with the first
step approximation, in which only the lowest energy eigenstate of
the standard pure pairing interaction part is taken into
consideration. Our analysis show that the standard pairing model
with the first step approximation displays similar pair structures
of first few exact low-lying states of the model, which, therefore,
provides a link between the two models.

\pacs{ 21.60.Cs, 21.60.Fw, 03.65.Fd,71.10.Li,74.20.Fg, 02.60.Cb}
\end{abstract}

\maketitle
\section{Introduction}
Pairing is an important residual interaction when a mean-field
approach is used as a starting approximation for a description of
nuclear structure. In particular, pairing correlations are essential
for a description of binding energies, odd-even effects,
single-particle occupancies, excitation spectra, electromagnetic
transition rates, beta-decay probabilities, transfer reaction
amplitudes, low-lying collective modes, level densities, and moments
of inertia, etc.$^{[1-3]}$ Two commonly used methods, the
Bardeen-Cooper-Schrieffer (BCS) \cite{[4]} and
Hartree-Fock-Bogolyubov (HFB) \cite{[2]} technique for finding
approximate solutions are well known. The limitations of these
methods, when applied in nuclear physics, are well
understood,$^{[5]}$ which is also the case when using these methods
to determine the energy spectra of nano-scale metallic
grains.$^{[6-7]}$ Various procedures have been used to correct the
approximation deficiencies, such as particle-number projected
mean-field treatments,$^{[8-10]}$ the use of coherent
states,$^{[11]}$ stochastic number projection techniques,$^{[12]}$
statistical descriptions,$^{[13]}$ treatments of the residual parts
of the Hamiltonian in the random phase approximation,$^{[14-15]}$
and various recursive approaches.$^{[16-17]}$  Typically, however,
these procedures have been found only limited applicability because
the results usually yield insufficient accuracy. Other methods are
likewise limited by their own sets of complications.

 On the other hand, an exact treatment of the nuclear mean-field plus
pairing type Hamiltonian was initiated by Richardson and Gaudin,
known as the Richardson-Gaudin method.$^{[18-21]}$ Recently,
extensions to the Richardson-Gaudin theory have been made using the
Bethe ansatz methodology,$^{[22-31]}$ especially, an application of
the Richardson-Gaudin solution to Sm isotopes with less than seven
pairs of valence nucleons was made.$^{[32]}$ Though these approaches
show that the mean-field plus pairing model is exactly solvable, the
solutions are generally not simple and normally require extensive
numerical work, especially when the number of levels and valence
pairs are large in spite of the recent efforts in improving the
procedure.$^{[33]}$ In [34], an extended pairing model with
many-pair interaction terms was proposed, which can be solved based
on a simpler Bethe ansatz, and describes even-odd mass differences
in $^{154-171}$Yb isotopes rather well. However, it was not clear
why the paring interaction strength in the model should be
drastically reduced with increasing of the number of valence nucleon
pairs in order to fit experimental data of even-odd mass
differences. Moreover, since the standard form of pairing
Hamiltonian is commonly adopted, it should be interesting to see
whether the extended paring model in some aspects covers main
features of the standard pairing model.

In the following, a new step-by-step diagonalization procedure for
evaluating exact solutions of the nuclear deformed mean-field plus
pairing interaction model will be proposed via a simple Bethe ansatz
in each step from which the eigenvalues and corresponding
eigenstates can be obtained progressively, which is shown in Sec.
II.  In Sec. III, as a demonstration of the procedure, a system with
$p=10$ levels will be analyzed. In Sec. IV, it will be shown that
the extended pairing model for deformed nuclei is similar to the
standard pairing model with the first step approximation, in which
only the lowest energy eigenstate of the standard pure pairing
interaction part is taken into consideration. Our analysis reveals
that the standard pairing Hamiltonian with the first step
approximation displays similar pair structures of first few exact
low-lying states of the model, which, therefore, provides a link
between the two models. A short discussion regarding implications of
our results is given in Sec. V.
\\\\
\section{ A new diagonalization procedure for the mean-field plus
standard pairing Hamiltonian}

The deformed mean-field plus standard pairing model Hamiltonian is
given by
\begin{equation}
\hat{H}= \sum_{j=1}^{p}\epsilon
_{j}\hat{n}_{j}-GS^{+}S^{-}=\sum_{j=1}^{p}\epsilon
_{j}\hat{n}_{j}-G\sum_{ij}S_{i}^{+}S^{-}_{j},
  \label{eqno1}
  \end{equation}
where $p$ is the total number of levels considered, $G>0$ is the
overall pairing strength, $\{\epsilon _{j}\}$ are single-particle
energies taken from any deformed mean-field, such as the Nilsson
model$^{[2]}$ or the relativistic mean-field theory,$^{[35-37]}$
$\hat{n}_{j}=c_{j\uparrow }^{\dagger }c_{j\uparrow}+c_{j\downarrow
}^{\dagger }c_{j\downarrow }$ is the fermion number operator for the
$j$-th level, and $S_{i}^{+}=c_{i\uparrow}^{\dagger }c_{i\downarrow
}^{\dagger }$ ($S^{-}_{i}=(S_{i}^{+})^{\dagger}=c_{i\downarrow
}c_{i\uparrow }$) are pair creation (annihilation) operators. The up
and down arrows in these expressions refer to time-reversed states.
Since the formalism for even-odd systems is similar, in the
following, we only focus on the even-even seniority zero case.

Since each deformed level can be occupied by no more than a single
pair due to the Pauli principle, Hamiltonian (1) is also equivalent
to a finite-site hard-core Bose-Hubbard model with infinite range
hopping and infinite on-site repulsion. Let $B^{+}_{i_{1}i_{2}\cdots
i_{k}}= S^{+}_{i_{1}}S^{+}_{i_{2}}\cdots S^{+}_{i_{k}}$ with $1\leq
i_{1} <i_{2}<\cdots<i_{k}\leq p$.  In the following, we set
$\mu=(i_{1}i_{2}\cdots i_{k})$ to be the $\mu$-th normal order
sequence with $1\leq i_{1} <i_{2}<\cdots<i_{k}\leq p$. The operator
$B^{+}_{\mu}$ can be regarded  as a grand hard-core boson creation
operator.$^{[34]}$  The total number of these operators is
$p!/((p-k)!k!)$ which exactly equals the dimension of the Hilbert
subspace of $k$ pairs with no double occupancy.

For $k$-pair excitations, by using the standard second quantization
formalism, the Hamiltonian (1) can effectively be reduced to the
following `one-body' Hamiltonian in the grand hard-core boson
picture:

\begin{equation}
\hat{H}_{k}=\sum_{\mu\nu} \langle\mu\vert \sum_{j=1}^{p}\epsilon
_{j}\hat{n}_{j} \vert\nu\rangle B^{+}_{\mu}B_{\nu}-G\sum_{\mu\nu}
\langle\mu\vert S^{+}S^{-}\vert\nu\rangle B^{+}_{\mu}B_{\nu},
\end{equation}
where $B_{\mu}=(B^{+}_{\mu})^{\dagger}$, because any $k$-pair
eigenstate of (1) can be expanded in terms of single grand hard-core
boson states $\{\vert\mu\rangle=B^{+}_{\mu} \vert 0\rangle\}$, where
$\vert 0\rangle$ is the pair vacuum state. In (2), the mean-field
one-body term in the Hilbert subspace spanned by
$\{\vert\mu\rangle=B^{+}_{\mu}\vert 0\rangle\}$ is diagonal with the
matrix elements

\begin{equation}
\langle\mu\vert \sum_{j=1}^{p}\epsilon _{j}\hat{n}_{j}
\vert\nu\rangle=\delta_{\mu\nu}2\bar{\epsilon}_{\mu}=
\delta_{\mu\nu}2\sum_{t=1}^{k}\epsilon_{i_{t}},
\end{equation}
while the matrix elements of the pairing interaction term are

\begin{equation}
\langle\mu\vert S^{+}S^{-}\vert\nu\rangle=\sum_{q\rho }
\langle\mu\vert p/2-q~\rho k\rangle_{\rm Q} {_{\rm Q}\langle} p/2-q~
\rho k\vert S^{+}S^{-}\vert p/2-q~\rho k\rangle_{\rm Q}~{_{\rm
Q}\langle} p/2-q~\rho k\vert\nu\rangle,\end{equation} in which

\begin{equation}
 _{\rm Q}\langle p/2-q~\rho k\vert S^{+}S^{-}\vert p/2-q~\rho
k\rangle_{\rm Q}= h^{(q)}_{k}=(k-q)(p-k-q+1)
\end{equation}
is matrix element of $S^{+}S^{-}$ in the Racah quasi-spin
formalism$^{[38]}$ with the total quasi-spin $S_{\rm Q}=p/2-q$,
where $q=0,1,\cdots, \min[k,p-k]$, and $\rho$ is an additional quantum number
needed in distinguishing from different states with  the same
quasi-spin for given $p$ and $k$. For given $p$, the total number of
different quasi-spin states with the same $S_{\rm Q}=p/2-q$ is given
by$^{[39]}$

\begin{equation}\omega_{q}={(p-2q+1)p!\over{(p-q+1)(p-q)!q!}}.
\end{equation}
 Furthermore, $\alpha^{q\rho}_{\mu}=\langle\mu\vert p/2-q~\rho k\rangle_{\rm Q}$ used
 in (4) is an overlap of the quasi-spin state
$\vert p/2-q~\rho k\rangle_{\rm Q}$ with $\mu$-th single grand
hard-core boson state $\vert\mu\rangle$, which can be chosen as
real. It can easily be verified that the total number of different
quasi-spin states equals exactly to the dimension of the Hilbert
subspace of $k$ pairs with no double occupancy:

\begin{equation}
d=\sum_{q=0}^{\min[k,p-k]}\omega_{q}={p!\over{(p-k)!k!}}.
\end{equation}

Thus, (2) can explicitly be written as

\begin{equation}
\hat{H}_{k}=\sum_{\mu=1}^{d}2\bar{\epsilon}_{\mu}B^{+}_{\mu}B_{\mu}-
G\sum_{q=0}^{\min[k,p-k]}h^{(q)}_{k}\sum_{\rho=1}^{\omega_{q}}
\sum_{\mu\nu}\alpha^{q\rho}_{\mu}\alpha^{q\rho}_{\nu}
B^{+}_{\mu}B_{\nu}.
\end{equation}

In order to diagonalize the Hamiltonian (8), let us consider a
simpler Hamiltonian  with only first term and $q=0$ part in the
second term of (8):

\begin{equation}
h_{0}=\sum_{\mu=1}^{d}2\bar{\epsilon}_{\mu}B^{+}_{\mu}B_{\mu}-
Gh^{(0)}_{k}\sum_{\mu\nu}\alpha^{01}_{\mu}\alpha^{01}_{\nu}
B^{+}_{\mu}B_{\nu}.
\end{equation}
As shown in [34, 40], The Hamiltonian (9) can be digonalized into
the following form:

\begin{equation}h_{0}=\sum_{\tau_{0}}E^{(\tau_{0})}
D^{+}(E^{(\tau_{0})})D(E^{(\tau_{0})})
\end{equation}
with

\begin{equation}
D^{+}(E^{(\tau_{0})})=\sqrt{1\over{{\cal
N}_{\tau_{0}}}}\sum_{\mu}{\alpha^{01}_{\mu}\over{2\bar{\epsilon}_{\mu}-E^{(\tau_{0})}}}
B^{+}_{\mu},
\end{equation}
where $E^{(\tau_{0})}$ is the $\tau_{0}$-th root of the following
equation:

\begin{equation}
Gh^{(0)}_{k}\sum_{\mu}{
(\alpha^{01}_{\mu})^2\over{2\bar{\epsilon}_{\mu}-E^{(\tau_{0})}}}=1,
\end{equation}
and ${\cal N}_{\tau_{0}}$ is the normalization constant obtained
from

\begin{equation}
\sum_{\mu}{(\alpha^{01}_{\mu})^{2}\over{
(E^{(\tau_{0})}-2\bar{\epsilon}_{\mu})
(E^{(\tau^{\prime}_{0})}-2\bar{\epsilon}_{\mu})
}}=\delta_{\tau_{0}\tau^{\prime}_{0}}{\cal N}_{\tau_{0}}.
\end{equation}
In this case, (12) will provide with
exactly $d$ different roots $E^{(\tau_{0})}$
as long as all combinations of the single-particle energies
$\bar{\epsilon}_{\mu}=\sum_{t=1}^{k}\epsilon _{i_{t}}$ are different for all $k$-pair
excitation cases. Fortunately, this is always the case when
single-particle energies $\{\epsilon_{j}\}$ are generated from any
deformed mean-field theory.

Since (8) and (9) should be diagonalized within the single grand
hardcore boson subspace spanned by $\{ B^{+}_{\mu}\vert 0\rangle\}$,
the effective commutation relations needed to prove that (9) can
indeed be expressed in the form shown in (10) are

\begin{equation}
[B_{\nu},B^{+}_{\mu}]=\delta_{\mu\nu},
\end{equation}
which are only valid when they are applied onto the vacuum state.
Using (9) and the ansatz (11), we have

\begin{equation}
[h_{0},D^{+}(E^{(\tau_{0})})]= E^{(\tau_{0})}D^{+}(E^{(\tau_{0})})-
\sqrt{1\over{{\cal N}_{\tau_{0}}}}(1-Gh^{(0)}_{k}\sum_{\mu}{
(\alpha^{01}_{\mu})^2\over{2\bar{\epsilon}_{\mu}-E^{(\tau_{0})}}})
\sum_{\nu}\alpha^{01}_{\nu}B^{+}_{\nu}.
\end{equation}
Though a direct proof is in demand,
it can be checked numerically with any set of parameters,
$Gh^{(0)}_{k}$, $\{\alpha_{\mu}^{01}\}$, and $\{\bar{\epsilon}_{\mu}\}$
with $\bar{\epsilon}_{1}\neq \bar{\epsilon}_{2}\cdots
\neq \bar{\epsilon}_{d}$, that the
orthnormal condition (13) is automatically satisfied when
$E^{(\tau_{0})}$ satisfies (12).
Therefore, (9) can indeed be expressed as that shown in (10) as long
as (12) is satisfied.

In order to simplify our expression, in the following,
the indices $(q,\rho)$ are relabeled by $r$ with $r=(q,\rho)$.
Thus, the Hamiltonian (8) can be rewritten as

\begin{equation}
\hat{H}_{k}=\sum_{\tau_{0}}E^{(\tau_{0})}
D^{+}(E^{(\tau_{0})})D(E^{(\tau_{0})})- G\sum_{r=1}^{d-1}h^{(r)}_{k}
\sum_{\mu\nu}\alpha^{r}_{\mu}\alpha^{r}_{\nu} B^{+}_{\mu}B_{\nu}.
\end{equation}

Then, we similarly have

$$
\sum_{\tau_{0}}E^{(\tau_{0})}
D^{+}(E^{(\tau_{0})})D(E^{(\tau_{0})})- Gh^{(1)}_{k}
\sum_{\mu\nu}\alpha^{1}_{\mu}\alpha^{1}_{\nu} B^{+}_{\mu}B_{\nu}=$$
\begin{equation}
\sum_{\tau_{0}}E^{(\tau_{0})}
D^{+}(E^{(\tau_{0})})D(E^{(\tau_{0})})- Gh^{(1)}_{k}
\sum_{\tau_{0}\tau^{\prime}_{0}} \Lambda_{\tau_{0}}
\Lambda_{\tau^{\prime}_{0}}
D^{+}(E^{(\tau_{0})})D(E^{(\tau^{\prime}_{0})})=
\sum_{\tau_{1}}E^{(\tau_{1})}
D^{+}(E^{(\tau_{1})})D(E^{(\tau_{1})}),
\end{equation}
where
\begin{equation}
\Lambda_{\tau_{0}}=
\sqrt{1\over{{\cal N}_{\tau_{0}}}}
\sum_{\mu}{\alpha_{\mu}^{01}\alpha^{1}_{\mu}\over{
2\bar{\epsilon}_{\mu}-E^{(\tau_{0})}
}},
\end{equation}

\begin{equation}
D^{+}(E^{(\tau_{1})})=\sqrt{1\over{{\cal N}_{\tau_{1}}}}
\sum_{\tau_{0}}{\Lambda_{\tau_{0}}\over{
E^{(\tau_{0})}-E^{(\tau_{1})}
}}D^{+}(E^{(\tau_{0})}),
\end{equation}
$E^{(\tau_{1})}$ is the $\tau_{1}$-th root of the following
equation:

\begin{equation}
Gh^{(1)}_{k}\sum_{\tau_{0}}{
(\Lambda_{\tau_{0}})^2\over{E^{(\tau_{0})}}-E^{(\tau_{1})}}=1,
\end{equation}
and ${\cal N}_{\tau_{1}}$ is the normalization constant obtained
from

\begin{equation}
\sum_{\tau_{0}}{(\Lambda_{\tau_{0}})^{2}\over{
(E^{(\tau_{0})}-E^{(\tau_{1})})
(E^{(\tau_{0})}-E^{(\tau^{\prime}_{1})})
}}=\delta_{\tau_{1}\tau^{\prime}_{1}}{\cal N}_{\tau_{1}}.
\end{equation}

Thus,using the results shown in (17)-(21)
and following the above procedure
consecutively,  we finally have

\begin{equation}
\hat{H}_{k}=\sum_{\tau_{0}}E^{(\tau_{0})}
D^{+}(E^{(\tau_{0})})D(E^{(\tau_{0})})- G\sum_{r=1}^{d-1}h^{(r)}_{k}
\sum_{\mu\nu}\alpha^{r}_{\mu}\alpha^{r}_{\nu} B^{+}_{\mu}B_{\nu}
=\sum_{\tau_{d-1}}E^{(\tau_{d-1})}
D^{+}(E^{(\tau_{d-1})})D(E^{(\tau_{d-1})}),
\end{equation}
where

\begin{equation}
D^{+}(E^{(\tau_{d-1})})= \sqrt{1\over{{\cal N}_{\tau_{d-1}}}}
\sum_{\tau_{d-2}}{\Lambda_{\tau_{d-2}}\over{
E^{(\tau_{d-2})}-E^{(\tau_{d-1})} }}D^{+}(E^{(\tau_{d-2})})
\end{equation}
with

\begin{equation}
\Lambda_{\tau_{s}}= \sqrt{1\over{{\cal N}_{\tau_{s}}}}
\sum_{\tau_{0}\tau_{1}\cdots\tau_{s-1}}
\prod_{\nu=0}^{s-1}{\Lambda_{\tau_{\nu}}\over{E^{(\tau_{\nu})}-E^{(\tau_{
\nu+1})} }} \sum_{\mu}{\alpha^{s+1}_{\mu}\alpha^{01}_{\mu}\over{
2\bar{\epsilon}_{\mu}-E^{\tau_{0}} }},
\end{equation}

\begin{equation}
D^{+}(E^{(\tau_{s+1})})= \sqrt{1\over{{\cal N}_{\tau_{s+1}}}}
\sum_{\tau_{s}}{\Lambda_{\tau_{s}}\over{
E^{(\tau_{s})}-E^{(\tau_{s+1})} }}D^{+}(E^{(\tau_{s})}),
\end{equation}

\begin{equation}
Gh^{(s+1)}_{k}\sum_{\tau_{s}}{
(\Lambda_{\tau_{s}})^2\over{E^{(\tau_{s})}}-E^{(\tau_{s+1})}}=1,
\end{equation}
and
\begin{equation}
\sum_{\tau_{s}}{(\Lambda_{\tau_{s}})^{2}\over{
(E^{(\tau_{s})}-E^{(\tau_{s+1})})
(E^{(\tau_{s})}-E^{(\tau^{\prime}_{s+1})})
}}=\delta_{\tau_{s+1}\tau^{\prime}_{s+1}}{\cal N}_{\tau_{s+1}},
\end{equation}
for $s=0,1,2,\cdots,d-2$. Hence, after $d$ steps, the Hamiltonian
(8) is diagonalized as shown in (22), of which eigenstate is

\begin{equation}
\vert k,\tau_{d-1}\rangle=D^{+}(E^{(\tau_{d-1})})\vert 0\rangle
\end{equation}
with the corresponding eigen-energy $E^{(\tau_{d-1})}$.

This new step-by-step diagonalization procedure needs at most $d$
steps to get final exact results, but in each step the corresponding
Bethe ansatz equation (26) contains only one variable, of which
roots can easily be obtained numerically similar to what is required
in the extended pairing model proposed previously,$^{[34]}$ and in
the TDA and RPA approximations with separable potentials.$^{[2]}$
Though this method may be unpractical for large size systems because
one needs to get all $d$ roots from the equation (26) in each step,
this procedure can also be used to check contributions from pairing
potential in the Racah quasi-spin formalism for different $q$ of the
second term in (8), and is certainly applicable to relatively small
systems. Actually, for $k$ pair excitation, though each term with
different $q$ from the second term of (8) will contribute to the
final eigen-energy and correlate with eigenstates, the first few of
these terms are key to determine properties of the first few
low-lying states of the model as will be shown in the next section.

\section{A numerical example for $\bf p=10$}

In this section, we will apply this new step-by-step diagonalization
procedure to the deformed mean-field plus  standard pairing model
for $p=10$ levels with number of pairs $k=1,2,\cdots, 10$,
in which the single particle energies
are given by $\epsilon_{i}=i+\chi_{i}$ for $i=1,2,\cdots,10$,
where $\chi_{i}$ are random numbers within the interval $(0,1)$
to avoid accidental degeneracy, and the pairing strength is set
to be $G=0.5$. since $h^{(0)}_{k}>h^{(1)}_{k}\geq\cdots\geq h^{d-1}$
is always satisfied, the lowest quasi-spin term with $q=0$
from the pairing potential should be most important to the first few
of eigenstates of the model, which is indeed the case as can be seen
from results shown in Table I.

In each step, we need to calculate the overlaps of the quasi-spin
states $\vert p/2-q~\rho k\rangle_{\rm Q}$ with $\mu$-th single
grand hard-core boson states $\vert\mu\rangle$. For $q=0$, the
$k$-pair state

\begin{equation}
\vert p/2~ k\rangle_{\rm Q}=\sqrt{(p-k)!/(k!p!)}(S^{+})^{k}\vert
0\rangle=
 \sqrt{k!(p-k)!/p!}\sum_{1\leq
i_{1}<i_{2}<\cdots<i_{k}\leq p}S^{+}_{i_{1}}S^{+}_{i_{2}}\cdots
S^{+}_{i_{k}}\vert 0\rangle= \sqrt{1/d}\sum_{\mu}B^{+}_{\mu}\vert
0\rangle
\end{equation}
is the eigen-state of the operator $S^{+}S^{-}$. Thus we have

\begin{equation}
\alpha^{01}_{\mu}=\langle\mu\vert p/2 k\rangle_{\rm Q} = \sqrt{1/d}.
\end{equation}
For $q\geq 1$,  the quasi-spin states $\vert p/2-q~\rho
k\rangle_{\rm Q}$ can be obtained by directly diagonalizing
$S^{+}S^{-}$ as shown in (5), or by using representation theory of
$SU_{\rm Q}(2)\times S_{p}$ summarized in [39]. Then, one can use
them to calculate the overlaps $\alpha^{q\rho}_{\mu}=\langle\mu\vert
p/2-q~\rho k\rangle_{\rm Q}$.

In Table I, we list first $5$ eigenenergies and overlaps of the
eigenstates with the corresponding exact ones for number of pairs
$k=1,2,\cdots, 10$ calculated with only $h^{0}$ term involved, which
is called the first step approximation. With the first step
approximation, it can be seen from Table I that results of $k=1$ and
$k=10$ cases are exact because $h^{(q)}=0$ for $q\geq 1$. The
approximate energy eigenvalues will gradually greater than the
corresponding exact ones with increasing of the number of pairs $k$
since pairing potential terms $h^{(q)}$ with
$q=1,2,\cdots,\min[p-k,k]$ will contribute more and more to the
final eigenenergies. However, the overlaps of the eigenstates with
the corresponding exact ones are always greater than $88\%$ for the
ground and the first excited states for any number of pairs $k$.
Therefore, $h^{(0)}$ term of the pairing potential is dominant in
determining pairing structure of the first two excitation states in
the model though the corresponding energy eigenvalues are different
from the exact ones.

\begin{table}
\caption{First $5$ eigenenergies of the standard pairing model with
$p=10$ levels for $k=0,1,2\cdots, 10$ obtained from the first step
approximation (appro.) and compared with the corresponding exact
results (exact), and overlap (olp) of the eigenstates obtained from
the first step approximation with the corresponding exact ones, in
which the single-particle energies $\epsilon _{1}=1.706$,
$\epsilon_{2}=2.754$, $\epsilon_{3}=3.440$, $\epsilon_{4}=4.349$,
$\epsilon_{5}=5.743$, $\epsilon_{6}=6.604$, $\epsilon_{7}=7.591$,
$\epsilon_{8}=8.959$, $\epsilon_{9}=9.335$, $\epsilon_{10}=10.125$,
and pairing strength $G=0.5$, where the single-particle energies and
$G$ are given in arbitrary units. }
\begin{tabular}{ccccccccccccccccc}
\hline \hline
        &&k=1&  &&k=2&   &&k=3&   &&k=4&     &&k=5&  \\
&exact&appro.&olp~~  &exact&appro.&olp~~ &exact&appro.&olp~~
&exact&appro.&olp~~
&exact&appro.&olp\\
        \hline
$E_{1}$ &2.255 &2.255  &100$\%$~~ &6.662&8.133&98$\%$~~
&12.873&15.423&93$\%$~~
&21.101 &24.272 &91$\%$~~ &31.856 &35.748 &88$\%$ \\
$E_{2}$ &4.797 &4.797  &100$\%$~~ &8.999&9.808&97$\%$~~
&15.508&17.213&92$\%$~~
&24.293 &26.909 &90$\%$~~ &34.529 &37.447 &88$\%$ \\
$E_{3}$ &6.438 &6.438  &100$\%$~~ &10.756&11.392&78$\%$~~
&16.961&18.576&89$\%$~~
&26.161 &28.312 &65$\%$~~ &36.407 &39.209 &83$\%$ \\
$E_{4}$ &8.272 &8.272  &100$\%$~~ &13.128&12.2618&75$\%$~~
&17.973&19.832&82$\%$~~
&26.176 &29.058 &67$\%$~~ &36.961 &40.167 &81$\%$ \\
$E_{5}$ &10.969 &10.969  &100$\%$~~ &13.375&13.625&85$\%$~~
&18.936&20.598&65$\%$~~
&27.621 &29.882 &62$\%$~~ &38.828 &41.206 &55$\%$ \\
\hline \hline
        &&k=6&  &&k=7&   &&k=8&   &&k=9&     &&k=10&  \\
&exact&appro.&olp~~  &exact&appro.&olp~~ &exact&appro.&olp~~
&exact&appro.&olp~~
&exact&appro.&olp\\
        \hline
$E_{1}$ &44.638 &48.868 &91$\%$~~&59.532 &63.713  &96$\%$~~
&76.780&79.982&96$\%$~~ &95.625&97.091&95$\%$~~
&116.212 &116.212 &100$\%$ \\
$E_{2}$ &47.415 &50.152  &90$\%$~~ &62.610&65.911&93$\%$~~
&78.949&82.641&97$\%$~~
&97.888 &101.652 &96$\%$~~ &&& \\
$E_{3}$ &49.165 &52.261  &81$\%$~~ &63.852&67.440&82$\%$~~
&80.277&83.861&81$\%$~~
&99.015 &102.961 &99$\%$~~ &&& \\
$E_{4}$ &50.908 &53.366  &76$\%$~~ &64.573&69.232&71$\%$~~
&81.182&85.191&82$\%$~~
&101.540 &105.221 &98$\%$~~ &&&\\
$E_{5}$ &51.728 &54.227  &74$\%$~~ &65.263&69.651&55$\%$~~
&83.148&86.404&75$\%$~~
&103.552 &107.341 &98$\%$~~ &&& \\
\hline
\end{tabular}
\end{table}

Since the largest deviation of the energy eigenvalues from the exact
ones occurs at the half-filling case, using the procedure shown in
the previous section, we calculated energy eigenvalues step by step
for $k=5$ case with $q=0,1,\cdots, 4$ since $h^{(5)}=0$, of which
the results are shown in Table II. For given $q$, there are actually
$\omega_{q}$ sub-steps involved in the diagonalization process
according to the procedure shown in the previous section. It can be
seen from Table II that the overlaps of the first $4$ eigenstates
with the corresponding exact ones will reach 99$\%$ after three
diagonalization steps though there is still deviation in
eigenenergies, which shows that the first few $h^{(q)}$ terms with
$q=0,1,2$ are key to determine pair structure of the first few
low-lying states of the model. While high-lying quasi-spin states
mainly correlate with high excited states of the model and keep low
part of the spectrum less affected.

\begin{table}
\caption{First $5$ eigenenergies of the standard pairing model with
$p=10$ levels for $k=5$ obtained from step-by-step diagonalization
procedure, in which the parameters used are the same as those shown
in Table I, where only $h^{(q)}$ with $q=0,1,2,\cdots, s$ terms from
the pairing potential are involved in the $s$-th step approximation
as described in the previous section. }
\begin{tabular}{ccccccccccccccccc}
\hline \hline
        &~~~1st step&  &~~~2nd step&   &~~~3rd step&
        &~~~4th step&     &~~~~exact  \\

 &~~~q=0&  &~~~q=0,1&   &~~~q=0,1,2&
        &~~~q=0,1,2,3&     &~~~~~~q=0,1,2,3,4\\
&eigenvalue&olp~~  &eigenvalue&olp~~ &eigenvalue&olp~~
&eigenvalue&olp~~
&eigenvalue&olp\\
        \hline
$E_{1}$ &35.748  &88$\%$~~ &33.862&98$\%$~~ &32.481&99$\%$~~
&32.176 &99$\%$~~ &31.856 &100$\%$ \\
$E_{2}$ &37.447  &88$\%$~~ &36.566&97$\%$~~&35.631&99$\%$~~
&35.102 &99$\%$~~  &34.528 &100$\%$ \\
$E_{3}$ &39.209  &83$\%$~~ &38.398&90$\%$~~ &37.451&99$\%$~~
&36.847 &99$\%$~~ &36.407 &100$\%$ \\
$E_{4}$ &40.167  &81$\%$~~ &39.422&89$\%$~~ &37.942&99$\%$~~
&37.637 &99$\%$~~ &36.961 &100$\%$ \\
$E_{5}$ &41.206  &55$\%$~~ &40.821&57$\%$~~ &39.779&97$\%$~~
&38.819 &97$\%$~~ &38.828 &100$\%$ \\
\hline \hline
 \end{tabular}
\end{table}

\section{comparison with the extended pairing model}

For $k$ pair excitations, if only $h^{(0)}$ term from the standard
pairing potential is considered for the standard pairing Hamiltonian
(8), namely

\begin{equation}
\hat{H}^{(1)}_{k}=\sum_{\mu=1}^{d}2\bar{\epsilon}_{\mu}B^{+}_{\mu}B_{\mu}-
G(k(p-k+1)/d)\sum_{\mu\nu} B^{+}_{\mu}B_{\nu},
\end{equation}
following (9)-(13), eigenstate of  (31) cen be written as
\begin{equation}
\vert
k;\tau)=\sum_{\mu=1}^{d}{1\over{2\bar{\epsilon}_{\mu}-E^{(\tau)}}}B^{+}_{\mu}\vert
0\rangle,
\end{equation}
where $\tau$ is an additional quantum number used to distinguish
different excitation states, and $E^{(\tau)}$ is an unknown variable
to be determined in diagonalizing (31). In solving the following
eigen-equation

\begin{equation}
\hat{H}^{(1)}_{k}\vert k;\tau)=E^{(\tau)}_{k}\vert k;\tau)
\end{equation}
with $B_{\mu}B^{+}_{\nu}\vert 0\rangle=\delta_{\mu\nu}\vert
0\rangle$, it shows that the variable $E^{(\tau)}$ must satisfy the
following equation:
\begin{equation}
G(k(p-k+1)/d)\sum_{\mu=1}^{d}{1\over{2\bar{\epsilon}_{\mu}-E^{(\tau)}}}=1,
\end{equation}
and the eigen-energy $E^{(\tau)}_{k}=E^{(\tau)}$. Thus, the
additional quantum number $\tau$ labels different roots of (34).
This is the so-called the first step approximation shown in section
II. The solution is complete so long as all combinations of the
single-particle energies $\sum_{t=1}^{k}\epsilon _{i_{t}}$ are
different for all $k$-pair excitation cases. Fortunately, this is
always the case when single-particle energies $\{\epsilon_{j}\}$ are
generated from any deformed mean-field theory. Since the single
grand particle energies $2\bar{\epsilon}_{\mu}$ are all different,
there are exactly ${p!/((p-k)!k!)}$ distinct roots in (34). The
resultant eigenstates (32), which are mutually orthogonal but not
normalized, satisfy

\begin{equation}
(k;\tau\vert k;\tau^{\prime})=\delta_{\tau\tau^{\prime}}{\cal
N}_{\tau},
\end{equation}
where

\begin{equation}
{\cal
N}_{\tau}=\sum_{\mu=1}^{d}{1\over{(2\bar{\epsilon}_{\mu}-E^{(\tau)})^{2}}}.
\end{equation}
It follows that the normalized eigenstate can be expressed as $\vert
k;\tau\rangle=\sqrt{1/{{\cal N}_{\tau}}}\vert k;\tau)$.

As shown in [34], a Nilsson mean-field plus extended pairing
interaction Hamiltonian

\begin{equation}
\hat{H}_{\rm ex}=\sum_{j=1}^{p}\epsilon _{j}n_{j}- G_{\rm
ex}\sum_{i,j}S_{i}^{+}S_{j} - G_{\rm ex}\left( \sum_{\rho
=2}^{\infty }{\frac{1}{{{(\rho !)}^{2}}}}\right.
\left.\sum_{i_{1}\neq i_{2}\neq \cdots \neq i_{2\rho
}}S_{i_{1}}^{+}S_{i_{2}}^{+}\cdots S_{i_{\rho}}^{+}S_{i_{\rho
+1}}S_{i_{\rho +2}}\cdots a_{i_{2\rho }}\right),
\end{equation}
where no pair of indices among $\{i_{1},i_{2},\cdots, i_{2\rho}\}$
are the same for any $\rho$, can also be solved exactly by using a
simple Bethe ansatz that is similar to what is proposed in this
work. Besides the usual Nilsson mean-field and the standard pairing
interaction, this form includes many-pair hopping terms that allow
nucleon pairs to simultaneously scatter (hop) between and among
different Nilsson levels. Furthermore, the extended pairing
interaction Hamiltonian (37) can be used to describe even-odd mass
differences rather well as long as the extended pairing interaction
strength $G_{\rm ex}$ decreases with an increasing number of pairs
k.  It follows from this that it is interesting to compare results
of the deformed mean-field plus standard pairing Hamiltonian (1)
with those from the extended pairing model.$^{[34]}$ And indeed, it
is not difficult to show that the expressions for eigenstates of the
extended pairing Hamiltonian and those of the standard pairing
Hamiltonian (1) in the first step approximation are the same. For
$k$-pair excitations, the eigenenergies $E^{(\tau)}_{k}(\rm ex)$ of
the extended pairing Hamiltonian (37) are given by

\begin{equation}
E^{(\tau)}_{k}(\rm ex)=E^{(\tau)}_{\rm ex}-(k-1)G_{\rm ex},
\end{equation}
where $E^{\tau}_{\rm ex}$ is the $\tau$-th root of the Bethe ansatz
equation,
\begin{equation}
G_{\rm
ex}\sum_{\mu=1}^{d}{1\over{2\bar{\epsilon}_{\mu}-E^{(\tau)}_{\rm
ex}}}=1.
\end{equation}
A comparison of (39) with the Bethe ansatz equation (32) for the
standard pairing Hamiltonian in the first step approximation (31)
shows that the two Hamiltonians yield exactly the same excitation
energies and the corresponding eigenstates so long as the parameter
$G_{\rm ex}$ in the extended pairing model and the parameter $G$ in
the standard pairing Hamiltonian (1) satisfy the following relation:

\begin{equation}
G_{\rm ex}=\left((p-k)!k!(p-k+1)k/p!\right)G.
\end{equation}
Furthermore, while the ground states of the two Hamiltonians are
also the same, the ground-state energies are different. However,
once the overall pairing strength $G$ is fixed, and the parameter
$G_{\rm ex}$ is chosen according to (40), it is easy to show that
the difference between the ground-state energy of the extended model
and that of the standard pairing model in the first step
approximation is given by

\begin{equation}
E^{({\rm g})}_{k}({\rm ex})-E^{({\rm g})}_{k}=-(k-1)G_{\rm ex}.
\end{equation}
This expression shows that the extended pairing interaction
contributes a little more attraction among valence pairs than the
standard pairing interaction in the first step approximation, but
reproduces excitation energies exactly the same as those in the
standard pairing model with first step approximation. Since $G_{\rm
ex}$ decreases drastically with increasing of $k$ toward the
half-filling, the ground sate energy difference of the two
Hamiltonians becomes negligible with increasing number of pairs $k$
with $k\leq [p/2]-1$ when $p$ is even, and $k\leq [p/2]$ when $p$ is
odd, where $[x]$ denotes integer part of $x$.

As an example, the ground-state energy difference (41) of the two
Hamiltonians in the sixth (82-126) major shell with the standard
pairing strength $G=0.2$MeV, which is a typical parameter value for
describing deformed nuclei in this region, shows that the
ground-state energy difference of the two Hamiltonians are rather
small in this case. The largest deviation of the ground-state energy
of the two Hamiltonians is at $k=2$ with $E^{({\rm g})}_{k}({\rm
ex})-E^{({\rm g})}_{k}=-36.3636$keV. Notwithstanding, since the only
difference between the two Hamiltonians, so long as $G_{\rm ex}$ is
taken to be related to $G$ by prescription (40), is in the overall
binding energy, and since an analytic expression for this difference
in also known in terms of $G_{\rm ex}$ through (41), for practical
purposes the two Hamiltonians yield the same results, even though
the Hamiltonians are quite different. This in itself is interesting,
since it shows that a many-pair interaction Hamiltonian can have
identical solutions to the two-pair interaction with truncations.
Obviously, it follows that for such systems the structure of fixed-Z
(isotopic) and fixed-N (isotonic) chains follow solely from the
structure the simplest single-pair member of the chain and simple
``pair-counting'' factors related to the pairing interaction
strength and single-particle energies.

Thus, we conclude that, basically, the extended pairing model is
different from the standard pairing model. However, if only a first
few eigenstates are considered, the pair structure of these states
in the two models are similar, especially in ground state, as can be
seen from analysis of the overlaps in the previous section. It can
be expected that the difference of the two models will be negligible
when number of pairs $k$ or pairing interaction strength $G$ is
small. In addition, since the extended pairing model can be solved
exactly with a single one variable equation (39), which is simpler
than the Richardson-Gaudin equations with $k$ variables for the
standard pairing Hamiltonian, the extended pairing model can be
applied to relative large systems, especially when one only want to
know a first few low-lying eigenstates and corresponding
eigenenergies.

\section{Conclusion}

A new step-by-step diagonalization procedure for evaluating exact
solutions of the nuclear deformed mean-field plus pairing
interaction model is proposed via a simple Bethe ansatz in each step
from which the eigenvalues and corresponding eigenstates can be
obtained progressively. This new approach draws upon an observation
that the original one- plus two-body problem in a $k$-particle
Hilbert subspace can be mapped unto a one-body grand hard-core boson
picture that can be solved step by step with a simple Bethe ansatz
known from earlier work, in which one only needs to solve a single
variable non-linear equation instead of a set of coupled non-linear
equations with $k$ variables as is required, for example, within the
framework of the well-known Richardson-Gaudin method. Though this
method may be unpractical for large size systems because one needs
to get all $d$ roots from the Bethe ansatz equation in each step,
this procedure can be used to check contributions from pairing
potential in the Racah quasi-spin formalism, and is certainly
applicable to relatively small systems.

As is shown in the example with $p=10$ levels, though each term with
different $q$ from the pure pairing interaction will contribute to
the final eigen-energy and correlate with eigenstates, a first few
of these terms are key to determine a first few low-lying states of
the model. While high-lying quasi-spin states mainly correlate with
high excited states of the model and keep low part of the spectrum
less affected.

Based on this new procedure, it is further shown that the extended
pairing model for deformed nuclei$^{[34]}$ is similar to the
standard pairing model with the first step approximation, in which
only the lowest energy eigenstate of the standard pure pairing
interaction part is taken into consideration. Our analysis show that
the standard pairing Hamiltonian with the first step approximation
displays similar pair structures of a first few low-lying states of
the standard pairing model, which, therefore, provides a link
between the two models.

Furthermore, the new method proposed is not limited to the deformed
mean-field plus pairing problem only, as it should also prove useful
for solving a much large class of quantum many-body problems in
which model Hamiltonians are described by
\begin{equation}
\hat{H}=\hat{H}_{0}+\lambda\hat{H}_{1},
\end{equation}
where $\lambda$ is a real parameter, and $\hat{H}_{0}$ and
$\hat{H}_{1}$ do not commute, $[\hat{H}_{0},\hat{H}_{1}]\neq0$.
According to our procedure, if the particle number is a conserved
quantity, and $\hat{H}_{0}$ and $\hat{H}_{1}$ can be diagonalized
independently in a $k$-particle basis, then the Hamiltonian (42) is
exactly solvable by using the step-by-step exact diagonalization
procedure. Moreover, the method can also be extended to deal with
Hamiltonians with more than two non-commutative terms by using a
similar procedure consecutively. Research in this direction is in
progress.

\section{Acknowledgements}

Support from the U.S. National Science Foundation (0500291), the
Southeastern Universities Research Association, the Natural Science
Foundation of China (10775064,~10575047), the Liaoning Education
Department Fund (20060464), and the LSU--LNNU joint research program
(9961) is acknowledged.

\end{document}